# Automated Detection of Configuration-Specific Security Vulnerabilities via Patch Analysis

FELIPE DE SANT'ANNA PAIXÃO, Federal University of Bahia, Brazil
JOANNA C. S. SANTOS, University of Notre Dame, USA
PAULO ANSELMO DA MOTA SILVEIRA NETO, Federal Rural University of Pernambuco, Brazil
DANIEL SADOC MENASCHE, Federal University of Rio de Janeiro, Brazil
GUSTAVO BITTENCOURT FIGUEIREDO, Federal University of Bahia, Brazil
EDUARDO SANTANA DE ALMEIDA, Federal University of Bahia, Brazil

We study how security patches in highly configurable C/C++ systems map onto the space of compile-time variants. We formalize the Vulnerability Impact Condition (VIC)—a Boolean predicate over configuration options that denotes all variants that contained the original flaw—and introduce *PatchLens*, a purely static technique that recovers VICs by aligning AST-level patch hunks with source-level presence conditions and resolving file inclusion via lightweight build system analysis. Evaluating *PatchLens* on 1,192 Linux kernel, 289 FFmpeg, and 100 PHP patches, we compute precise, human-readable VICs without the need to compile any system variant. The resulting predicates are compact (avg. 1.84 variables for Linux, 3.23 for FFmpeg, 1.04 for PHP) and show that only a small fraction of vulnerabilities are system-wide, which carry higher CVSS scores; meanwhile, CVE texts almost never encode the required options (≈1% average recall), motivating automated enrichment of CVE descriptions with VICs. *PatchLens* and the accompanying dataset enable immediate applications in CI (variant-aware triage and test selection), targeted sampling and fuzzing, and feature risk scoring, offering a scalable, explainable path to vulnerability assessment in highly configurable software.

CCS Concepts: • **Security and privacy** → **Vulnerability management**.

Additional Key Words and Phrases: Patch, vulnerability, CVSS, highly configurable systems

## 1 Introduction

Security vulnerabilities have proliferated in the last two decades, climbing from 2,707 disclosed flaws in 2004 to 8,982 in 2014, and reaching 38,317 in 2024 [8]. These vulnerabilities undermine critical security properties, enabling data breaches, unauthorized access, and denial-of-service attacks, often with substantial financial repercussions [13, 29, 46, 55].

Authors' Contact Information: Felipe de Sant'Anna Paixão, felipe.paixao@ufba.br, Federal University of Bahia, Salvador, Bahia, Brazil; Joanna C. S. Santos, joannacss@nd.edu, University of Notre Dame, Notre Dame, Indiana, USA; Paulo Anselmo da Mota Silveira Neto, paulo.motant@ufrpe.br, Federal Rural University of Pernambuco, Recife, Pernambuco, Brazil; Daniel Sadoc Menasche, sadoc@ufrj.br, Federal University of Rio de Janeiro, Rio de Janeiro, Rio de Janeiro, Brazil; Gustavo Bittencourt Figueiredo, gustavobf@ufba.br, Federal University of Bahia, Salvador, Bahia, Brazil; Eduardo Santana de Almeida, eduardo.almeida@ufba.br, Federal University of Bahia, Salvador, Bahia, Brazil.

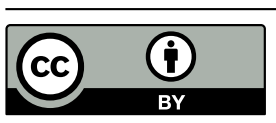

From a security perspective, the ever-growing complexity of configurable systems inherently creates complex challenges in testing, analysis, and maintenance. Many foundational software systems, such as the Linux kernel, OpenSSL, and system libraries, employ compile-time variability via feature flags or configuration options. Conditional macros in source code and rules in build scripts determine which code fragments are included in each build, inevitably leading to an exponential number of possible variants in relation to the number of configuration options [15, 34, 36, 39].

This variability poses a fundamental question: *how do security vulnerabilities relate to compile-time configurations?* Most existing analysis and refactoring tools assume a single fixed configuration, and therefore cannot scale to the combinatorial complexity of a highly configurable system [15, 18]. Although variability-aware analysis techniques have been proposed [12, 18, 27, 28, 34], their applicability to vulnerability detection and patch-impact assessment remains largely unexamined.

Previous work has explored the use of patch and commit analysis to identify Vulnerability-Contributing Commits (VCCs) in highly configurable systems [4, 47]. In contrast, we reason about the relationship between a security patch and the set of variants that remain vulnerable: while a patch may not affect every possible variant of a highly configurable system, it must encompass all variants that contained the original vulnerability. Consequently, by computing the entire subspace of variants impacted by a given security patch - which we call the *Vulnerability Impact Condition* (VIC) – one directly obtains a superset of all vulnerable variants, obviating the need for extensive variant testing.

A direct way to address this question is to inspect the code modified by security patches. Landsberg et al. [32] introduced SiB ("Should I Bother?"), a solution which hooks into Clang to filter out unaffected variants by matching patch hunks to variant-specific line ranges. However, SiB is only capable of evaluating the impact on pre-built variants and cannot produce a unified VIC across the entire configuration space. In contrast, the full *Vulnerability Impact Condition* of a patch — a Boolean formula over the configuration options — reduces variant filtering to a simple expression evaluation, eliminating the need for per-variant builds.

For example, consider CVE-2015-1421, a Linux-kernel vulnerability assigned a critical CVSS base score of 10.0. The flaw is a use-after-free in SCTP handling, which can be triggered remotely to cause a kernel panic (and potentially other impact). At face value, such a score suggests an urgent, system-wide patch rollout. However, the vulnerable code is only reachable in variants compiled with `CONFIG_IP_SCTP`, the option that enables Stream Control Transmission Protocol (SCTP) support—and this option is *not enabled by default* [1]. As a result, many systems running affected kernel versions are not vulnerable in practice simply because SCTP support is absent in their build. Crucially, this compile-time requirement is not reflected in the CVE text, which describes the bug but does not mention `CONFIG_IP_SCTP`. PatchLens makes this gap explicit by extracting a Vulnerability Impact Condition (VIC) from the security patch—here, essentially the predicate `CONFIG_IP_SCTP` (and the corresponding file-inclusion constraints)—so operators can decide exposure by checking a single configuration flag rather than treating all builds of a version as equally affected. This enables targeted patch rollouts, configuration-based mitigations (e.g., disabling the implicated feature where feasible), and enriched vulnerability advisories with precise compile-time requirements, among other significant impacts discussed in Section 5.

Motivated by this gap, we explore the extent to which static analysis of source code not processed yet by C's preprocessor and build system scripts can produce such impact conditions. We present *PatchLens*, a tool that parses patch diffs at the Abstract Syntax Tree (AST) level, computes presence conditions for all modified parts of the code, and resolves build system variability to inject accurate file inclusion predicates. Our evaluation demonstrates that *PatchLens* computes precise and human-readable impact conditions for thousands of patches in large C/C++ ecosystems, such as the Linux kernel, FFmpeg, and PHP, without requiring variant compilation. This capability opens new avenues for continuous integration (CI), targeted patch filtering strategies, and security-focused variant sampling.

[1] In the default `defconfigs` for the x86-64, x86, ARM, ARM64, and RISC-V at the time of this research.

Our study is guided by the following research questions:

- **RQ1:** ***To what extent can PatchLens statically analyze real-world vulnerability patches, and what is the accuracy of the results?*** To answer RQ1, we applied *PatchLens* to 1,192 Linux kernel patches, 289 FFmpeg patches, and 100 PHP patches, and recorded for each whether a presence-condition was successfully computed or, if not, whether the failure was due to non-C artifacts, parsing/build-system errors, or missing commits. Then, we manually reviewed a significant amount of vulnerability impact conditions, confirming their accuracy.
- **RQ2:** ***To what extent do security vulnerabilities impact all variants of a statically configured system, and are broadly-impacting vulnerabilities rated as more severe?*** To answer RQ2, we applied *PatchLens* to a dataset of patches of three systems, then computed each patch's presence-condition formula, and identified those that simplify to a tautology. We then retrieved the corresponding NVD Common Vulnerability Scoring System (CVSS) base scores and compared these scores of system-wide vulnerabilities against the global average.
- **RQ3:** ***How do vulnerability impact conditions look like?***
  - **RQ3.1:** ***On average, how many configuration variables appear in each presence-condition formula?***
  - **RQ3.2:** ***Which configuration options are most frequently implicated?***
  - **RQ3.3:** ***What are the most frequent binary operators (e.g., conjunction $\wedge$, disjunction $\vee$) used in these formulas?***

  To answer RQ3.1 and RQ3.2, we parsed every impact formula produced by *PatchLens* to count distinct configuration variables per patch and aggregated frequency counts to produce top-10 option lists for Linux Kernel, FFmpeg, and PHP (Table 3). To answer RQ3.3, we analyzed the same set of formulas to compute occurrences of each binary operator, thereby identifying which logical connectives dominate vulnerability impact conditions.
- **RQ4:** ***How much information can PatchLens add to CVE descriptions?*** To answer RQ4, we constructed a fuzzy-matching pipeline over 1,405 successfully analyzed CVEs. For each, we extracted the ground-truth option set $G_i$ from its impact formula, tokenized the NVD description, and used Python's `difflib` to detect mentions of these options. We then computed the coverage rate and average recall of these options.

Overall, the contributions of this paper are threefold:

- *PatchLens*: an automated and build system agnostic approach to analyze a patch to a HCS and identify the affected variants;
- Implementation of build system specific analysis solutions to determine the presence condition of C source files inside Linux, PHP, and FFmpeg;
- A large-scale dataset of security patches and their affected configuration in three large, highly configurable systems (Linux Kernel, PHP, and FFmpeg).

All source code, data from experiments, and instructions on how to replicate this study can be found in the supplementary materials [44] and, for future updates, in our public repository [43].

## 2 Background

This section presents the main concepts involved in this work. We first review static variability mechanisms in highly configurable C/C++ systems, the role of build systems in feature selection, the cataloging of vulnerabilities in CVE databases, and the unified-diff format used for security patches.

## 2.1 Highly Configurable Systems (HCS)

**Software variability** is a fundamental aspect of highly configurable systems (HCS) and software product lines (SPLs), allowing for the derivation of customized variants from a common codebase [3, 24, 25, 31, 45]. The Linux kernel is a well-known example of a highly configurable software system containing more than ten thousand configuration options [1, 12, 24, 25, 36, 50, 51]. The primary motivation for structuring systems as HCS is to allow tailoring for specific purposes and needs while minimizing code redundancy [25]. However, managing such extensive variability introduces significant complexity and poses challenges for traditional software engineering tools and practices [1, 12, 36, 37, 50, 51]. Concretely, the combinatorial explosion of configurations renders exhaustive testing infeasible, necessitating variant *sampling* rather than complete coverage [37]—and variability is often dispersed between source code and build systems, whose heterogeneous mechanisms and constraints complicate analysis, maintenance, and prevention of subtle features-interaction faults [1, 12, 51].

*2.1.1 Variability Implementation.* Highly configurable systems (HCS) may realize variability either at *compile time* (statically) or at *run time* (dynamically) [15, 40]. In statically configured software, the set of *configuration options* (features) is resolved before or during compilation, so that the actual source code fed to the compiler varies with the chosen options [34]. This work focuses exclusively on such *compile-time* variability. We do so because, in C/C++ ecosystems, source code variability is heavily implemented with compile-time and build-time options [12, 28, 34], making this layer decisive for exposure analysis.

In C and C++ projects, static variability is most commonly implemented via preprocessor directives [15, 27, 36]. Developers wrap code fragments—declarations, definitions, statements, or entire files—in boolean-valued macros using constructs such as those shown in Listing 1.

Listing 1. Variability via preprocessor macros

```
1 #if defined(FEATURE_A) && !defined(FEATURE_B)
2    /* included only when FEATURE_A is enabled
3       and FEATURE_B is disabled */
4 #endif
5 #ifdef FEATURE_C
6    void foo() { ... }
7 #endif
```

These `#if`, `#ifdef`, `#ifndef`, `#elif`, and `#endif` directives produce a distinct Abstract Syntax Tree (AST) for each assignment of feature macros. A *configuration* is a total assignment of truth values to the feature set, and a *variant* is the product obtained under that configuration. Without constraints, $n$ binary options admit $2^n$ variants, leading to a combinatorial explosion.

Despite its ubiquity, the C preprocessor is criticized for error-proneness and for scattering and tangling feature code, which hinders comprehension and maintenance [14, 18, 27, 34, 38, 39]. Macros further amplify complexity through code transformations and intricate interactions with conditional compilation [14, 18, 27, 38]. Empirical studies of annotation patterns show that although many `#ifdef` regions align with logical code boundaries, a significant fraction do not, complicating automated tool support [34, 35, 38]. Consequently, there is a growing demand for analysis and transformation tools capable of handling unpreprocessed source code directly [14, 15, 18, 27, 32, 34, 38, 39].

Additionally, HCS typically employs a *Feature Model* [26], which defines both the hierarchical organization of features and cross-tree constraints. In the hierarchy, a feature may be marked as *mandatory*, in which case it is automatically included whenever its parent is selected, or *optional*, allowing independent inclusion. Sibling features can be grouped into sets *alternative* (XOR), where exactly one member must be chosen, and sets *or* (OR), where at least one member must be chosen when the parent feature is enabled. Beyond the hierarchy, the constraints *requires* express that the selection of one feature implies another (e.g. $\text{FEATURE}_B \implies \text{FEATURE}_A$), while the constraints *excludes* prohibit certain combinations of features. Feature models are commonly translated into propositional formulas and analyzed with SAT solvers to ensure that only semantically valid configurations are considered [5, 6, 41].

In this context, a *presence condition* is a Boolean formula over configuration options that denotes the subset of variants in which a particular code artifact—such as a function or module—is included [1, 49, 52]. Presence conditions provide the formal link between feature selections and code inclusion semantics, forming the foundation for tools like *PatchLens*. Indeed, given a patch, *PatchLens* produces as output a vulnerability impact condition, which is a presence condition for that patch and, consequently, the vulnerability that it fixes.

*2.1.2 Build systems for HCS.* A *build system* orchestrates the transformation of source files into executable artifacts by specifying targets, dependencies, and the commands needed to produce each target. In HCS, variability can be realized not only via preprocessor directives in the source code but also through build-system logic. For instance, the Linux kernel employs both conditional macros (`#ifdef`/`#if`) and Kbuild/Makefile rules to select which files are compiled or which modules are linked [12].

GNU Make, the core of many C/C++ build processes, is effectively a small programming language supporting variable assignment and expansion, conditional directives (`ifeq`, `ifdef`), pattern rules, functions for string/manipulation, and "shell escapes" that execute arbitrary commands [12]. Such features, while powerful, render static analysis of build-time variability especially challenging: the presence condition of a source file may depend on nested variable expansions, recursive `include` directives, or the output of runtime shell commands [2, 12].

Empirical studies of the Linux kernel indicate that over 65% of feature-selection logic resides within the build system rather than the C source itself [12]. Consequently, any sound variability analysis must incorporate build system semantics to avoid under-approximating the set of valid variants. Additionally, the Linux kernel build system supports *tristate* configuration options, which select features to be compiled and bundled inside the kernel image itself, to be compiled and bundled separately as kernel modules, or to not be compiled at all [10].

However, the lack of a standardized build framework in C/C++ – with projects adopting Make[2], CMake[3], Bazel[4], Meson[5], custom scripts, or combinations thereof – means that build system analysis often becomes project-specific. A comprehensive tool must therefore support multiple build system languages natively or provide a modular architecture for integrating custom parsers and evaluators. In *PatchLens*, we record the file-level variability of the build system as special variables during AST analysis and then resolve it with lightweight project-specific build resolvers (Linux/Kbuild, PHP/Autotools, FFmpeg/Make) to obtain file inclusion predicates; see Section 3.3 for details.

[2]https://www.gnu.org/software/make/
[3]https://cmake.org/
[4]https://bazel.build/
[5]https://mesonbuild.com/

## 2.2 Software Vulnerabilities

Software vulnerabilities are software defects that result in violations of system security properties, potentially enabling unauthorized actions. Such vulnerabilities can have severe consequences, including unauthorized access, leakage of sensitive data, or denial-of-service attacks [13, 29].

Typically, vulnerabilities are publicly disclosed and systematically cataloged in specialized databases using a standardized format known as Common Vulnerabilities and Exposures (CVE). A prominent example is the National Vulnerability Database (NVD), which maintains comprehensive records of publicly disclosed vulnerabilities, currently cataloging around 300,000 CVEs [6]. Each CVE entry in the NVD database includes detailed metadata, such as a textual description, impact and severity metrics (*e.g.*, Common Vulnerability Scoring System - CVSS scores), references to external resources, weakness classifications (*e.g.*, , improper authentication, uninitialized pointer access), and affected configurations. Figure 1 shows an example of the structure of a typical CVE entry provided by the NVD.

Example of CVE Entry

**CVE-ID:** CVE-2023-1513
**Description:** A flaw was found in KVM. When calling the `KVM_GET_DEBUGREGS ioctl`, on `32-bit` systems, there might be some uninitialized portions of the `kvm_debugregs` structure that could be copied to userspace, causing an information leak.
**References:**

- URL to issue tracker
- URL to patch commit
- `[...]`

**Configurations:**

- `cpe:2.3:o:linux:linux_kernel:*:*:*:*:*:*:*:*` (up to 6.2)

Fig. 1. Example of a CVE Entry in the National Vulnerability Database

However, it is important to note that the configuration information provided by NVD does not explicitly address compile-time configuration options typical of highly configurable systems. Instead, it identifies vulnerable platforms using the Common Platform Enumeration (CPE), describing the combination of software products and specific affected versions [42]. In such context, *PatchLens* can be used to enrich CVE descriptions by explicitly linking them to the compile-time feature selections that expose specific vulnerabilities, and leveraging patches for that matter.

## 2.3 Vulnerability Patches

Software vulnerabilities are mitigated by edits to the source code, delivered as *patches*. In practice, most patches use the *unified diff* format (also known as the `unidiff` format), which encodes one or more file-level modifications as sequences of added and removed lines [21]. Within each file's diff, changes are grouped into *hunks*: contiguous blocks of line insertions and deletions that highlight the context around each modification.

A `unidiff` patch thus consists of one or more file diffs, each containing a sequence of hunks. For example, Listing 2 shows the content of the patch that fixes the vulnerability described by the CVE-2023-1513 [7] at Figure 1.

Although essential for security maintenance, vulnerability patches are not trivial to collect: they are dispersed across issue trackers, mailing lists, vendor advisories, and various online archives [53].

[6] In July, 2025. Information available at https://nvd.nist.gov/general/nvd-dashboard
[7] https://nvd.nist.gov/vuln/detail/cve-2023-1513

Listing 2. Contents of the patch that fixes CVE-2023-1513

```
1  --- a/arch/x86/kvm/x86.c
2  +++ b/arch/x86/kvm/x86.c
3  @@ -5263,12 +5263,11 @@ static void kvm_vcpu_ioctl_x86_get_debugregs(struct kvm_vcpu *vcpu,
4  {
5    unsigned long val;
6
7  + memset(dbgregs, 0, sizeof(*dbgregs));
8    memcpy(dbgregs->db, vcpu->arch.db, sizeof(vcpu->arch.db));
9    kvm_get_dr(vcpu, 6, &val);
10   dbgregs->dr6 = val;
11   dbgregs->dr7 = vcpu->arch.dr7;
12 - dbgregs->flags = 0;
13 - memset(&dbgregs->reserved, 0, sizeof(dbgregs->reserved));
14 }
```

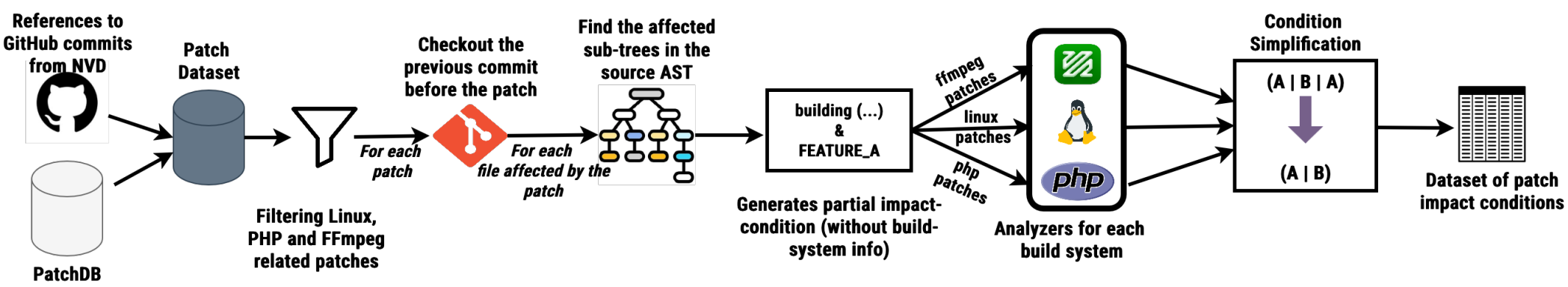

Fig. 2. Steps taken to generate the dataset with patches and their impact condition.

To quantify this dispersion, we analyzed the NVD 2.0 JSON feeds [8] from 2002 through July 2025. Of the 299,940 CVEs indexed, only 76,096 (≈ 25.37%) include at least one reference to a patch, and a mere 12,392 [9] CVEs reference a GitHub commit from which a patch file can be directly obtained.

## 3 Research Design

To evaluate the efficacy and performance of *PatchLens*, we follow a structured workflow that encompasses data collection, tool implementation, and impact-condition generation. First, we assemble a comprehensive corpus of security patches. Next, we describe the static analysis pipeline of *PatchLens*, including AST-based presence-condition extraction, build system variability resolution, and Boolean simplification. Finally, we apply the complete toolchain to our dataset, validate, and analyze the resulting impact formulas. The end-to-end process to generate the dataset is summarized in Figure 2. Detailed results are reported in Section 4.

### 3.1 Patch Dataset Preparation

To evaluate *PatchLens* on real-world security fixes, we first assembled a corpus of vulnerability patches. We used NVD's 2.0 JSON feeds, retrieving all CVEs from 2002 through July 2025. Each CVE entry includes a list of *references*, some of which point to patch artifacts; however, many references correspond to mailing-list threads, advisories, or project webpages rather than direct patch files. Thus, to obtain a reliably downloadable patch set, we filtered the NVD references for URLs matching GitHub commit patterns (*i.e.*, URLs of the form *https://github.com/<repo>/commit/<hash>*). Then, we retrieved the corresponding patch files from these URLs. This procedure yielded 12 392 distinct patches.

[8]Available at https://nvd.nist.gov/vuln/data-feeds
[9]This number may vary, as CVE information is continuously updated by the NVD. This measurement was done in June 2025.

To broaden coverage, we then integrated the PatchDB dataset [53], which provides 4 076 CVE-associated patches collected from diverse sources. After removing duplicate records against our GitHub-derived set, the combined corpus comprises 14 897 patches drawn from approximately 4 000 distinct software projects. We note that these figures may change over time as NVD reference metadata is updated.

Because *PatchLens* operates on unpreprocessed C/C++ and relies on resolving both pre- and post-patch revisions, we applied two conservative filters to the raw corpus before analysis. First, we excluded patches whose diffs modify no C translation units (e.g., only scripts, tests, build files, or documentation), since *PatchLens* cannot derive presence conditions for non-C artifacts. Second, we discarded patches whose referenced commits could not be resolved in the target repository (e.g., missing or rebased history), as computing AST-level changes requires checking out the immediate predecessor of each patched file. These filters reduce noise and prevent unsound inferences, while aligning the dataset with the technical scope of our analysis pipeline (see Section 4 for resulting coverage and failure modes). We note that this choice may underrepresent vulnerabilities confined to non-C assets, which we discuss as a threat to validity in Section 6.

In this study, we chose to further analyze Linux, FFmpeg, and PHP—highly configurable C codebases—because they rank among the most frequently patched macro/build-variability systems in our corpus (Linux #1: 1,483; FFmpeg #3: 298; PHP #6: 144) and jointly maximize diversity in both system domain and build-system design (Linux Kbuild/Kconfig, FFmpeg configure + GNU Make, PHP Autotools).

### 3.2 PatchLens

*PatchLens* is a purely *static* analysis tool for determining exactly which variants in a HCS are affected by a given vulnerability patch. Unlike approaches that require building or testing configurations, *PatchLens* operates directly on "*unpreprocessed*" source code (*i.e.*, code not processed yet by C's preprocessor) and patch diffs, producing a human-readable boolean expression that precisely characterizes the impacted subspace of variants.

*PatchLens* takes as input *(i)* a patch in the output format of `git format-patch`, and *(ii)* the path to the root of the software's source repository. Then, it analyzes each file-level diff in the patch to extract the modified code regions, or "hunks", parses the target file *without preprocessing* to obtain ASTs for both the pre- and post-patch versions, and maps each hunk to its corresponding AST nodes by comparing these two trees.

Each hunk corresponds to a modification of a subtree in the AST of the affected file. For every root node of the modified subtrees, *PatchLens* computes its *presence condition* by:

(1) Locating the AST node corresponding to the hunk's root.
(2) Traversing upward from the hunk's root node to the translation-unit root, collecting all variability conditions imposed by surrounding preprocessor directives. We used the same preprocessor conditional directives that are used by TypeChef [28]: `#ifdef`, `#ifndef`, `#if`, `#elif`, `#endif`. In this step, *PatchLens* will ignore conditions that match `~([...]_H)` or `~([...]_H_)` to filter out header guards [10].
(3) Conjoining these conditions yields the exact predicate under which that subtree is included in a variant.

For parsing, because *PatchLens* primarily relies on AST nodes associated with preprocessor directives, we use a custom, minimal parser built with Tree-sitter [33]. Prior work such as SuperC and TypeChef pioneered variability-aware parsing of *unpreprocessed* C into rich AST representations [18, 28]; however, because *PatchLens* does not require a complete variability-aware AST to determine presence conditions, a lightweight parser is sufficient in our setting. This strategy reduces parser complexity and, in turn, decreases the likelihood of failures compared to full-featured parsers for *unpreprocessed* C. Moreover, Tree-sitter provides error-tolerant parsing; when combined with our hunk-to-subtree strategy, *PatchLens* can still approximate presence conditions in the presence of local parse errors. Concretely, if the parser cannot reliably interpret a particular conditional directive (e.g., an `#ifdef` construct), *PatchLens* detects the issue and falls back to a more conservative (i.e., more general) condition. This design decision from *PatchLens* yields false positives (VICs that may include safe variants) rather than false negatives (omitting vulnerable variants), which is the preferred failure mode for security triage. Illustrative examples of this behavior are included in our supplementary materials [44].

[10]Header guards are preprocessor macros that prevent a header file from being included multiple times by wrapping its contents in an `#ifndef/#define/#endif` block.

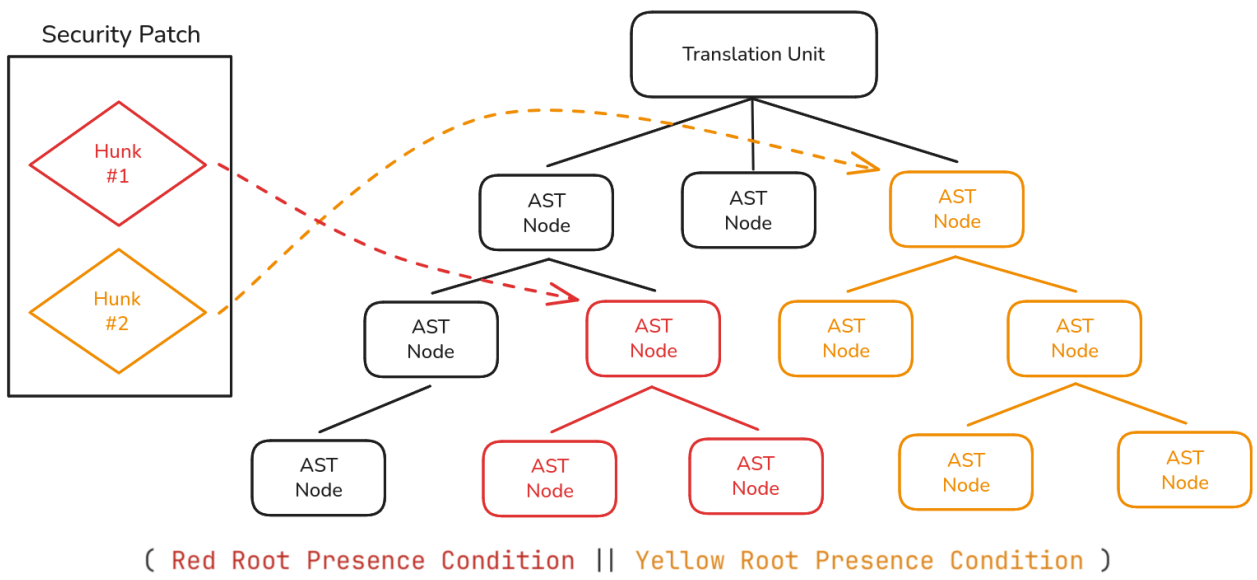

Fig. 3. Illustrative example of how *PatchLens* calculates the impact condition of a patch from the affected nodes in the AST.

Finally, *PatchLens* aggregates the presence conditions of all the hunks with a disjunction: a variant is 'affected' if *at least one* of the modified subtrees is included in that variant. Figure 3 illustrates how *PatchLens* computes a file's impact condition: it maps each patch hunk to its corresponding AST subtree, derives the presence condition at each subtree root by conjoining the enclosing variability guards, and then *disjoins* the resulting conditions across hunks to obtain the final predicate.

### 3.3 Build System Variability

The initial analysis of *PatchLens* produces a partial impact formula that captures only *source-level* presence conditions. To signal file-inclusion conditions, it injects a special function into the formula (*i.e.*, `building("path/to/file")`), to be expanded later into the actual variability conditions implemented at build-system level. Build-system analysis is necessary because many projects implement variability not only via conditional macros in the source code but also at the build-system level [2]. For example, the Linux kernel uses Kbuild/Makefile files to select which source files are compiled and how they are linked into the final image.

Determining the true presence condition of a file is non-trivial, since C and C++ lack a standardized build system: each project may adopt a different mechanism (Make, CMake, Bazel, *etc.*). Consequently, part of the *PatchLens* analysis process is specific to the build system. The extraction of the impact condition is divided into two stages. First, variability is extracted from the source code alone, agnostic to the build system, yielding expressions such as:

```
(CONFIG_MMU && building("mm/mmap.c"))
```

where `CONFIG_MMU` arises exclusively from `#ifdef` macros, and `building("mm/mmap.c")` stands for the file inclusion predicate. Next, a build-system-specific component is responsible for resolving each `building()` call by parsing the project's build scripts to compute the actual file-presence condition.

*Build System Analysis for Linux.* To resolve `building()` predicates in Linux, *PatchLens* incorporates a build-system resolver inspired by KBuildMiner [7]. KBuildMiner performs a global, recursive traversal of all Kbuild and Makefiles, constructs an in-memory model of variable assignments and build targets, and derives a presence condition for every C source file. Although comprehensive, this end-to-end analysis is inefficient for vulnerability patches—which typically modify only a handful of files[11]—and would require recomputing conditions for every kernel version.

Instead, our resolver inverts the process. Given a *target* C file, it:

(1) locates the nearest `Kbuild/Makefile` in the file's directory,
(2) parses that file into a tree of build rules and variable definitions,
(3) extracts the local inclusion predicate for the target (e.g., the guard that produces `core.o`),
(4) ascends to the parent directory and repeats steps 1–3, and
(5) conjoins all directory-level predicates to obtain the file's overall presence condition.

For example, consider deriving the presence condition for `/net/nfc/core.c`. The resolver first inspects `/net/nfc/Kbuild` to identify the rule that produces `core.o` and records its guard (e.g., a predicate over a feature such as `CONFIG_NFC`). It then moves up to `/net/Kbuild` to capture the predicate that governs inclusion of the `nfc/` subdirectory (e.g., a networking-level condition), and finally consults the top-level `Kbuild/Makefile` for any root-level guards. The resulting file presence condition is the conjunction of these directory predicates:

$$P_{\text{root}} \ \wedge \ P_{\text{net}} \ \wedge \ P_{\text{nfc}}.$$

This is precisely the predicate used to expand `building("net/nfc/core.c")` in the patch's impact formula.

When the resolver encounters nonstandard constructs (e.g., shell escapes or custom functions) that prevent sound interpretation, it marks the file's condition as *unresolved* and reports a resolver failure rather than guessing inclusion or exclusion. Per-project success rates are reported in Section 4. In addition, to account for the kernel's tristate options ($y/m/n$), we coerce $y$ and $m$ to `true`: both built-in and modular selections compile the target's code and therefore include it in the produced build artifacts.

*Build system analysis for PHP.* PHP is a popular open-source server-side scripting language designed for web development. Building PHP relies on the Autotools suite, especially Autoconf and Automake, which processes M4-based files through multiple generator stages (M4 preprocessing, `configure` script generation, template substitution, and Make execution) to produce build artifacts executed in later stages [2, 20]. This multistage, cross-language workflow is notoriously difficult to comprehend: developers have nicknamed it "auto-hell" due to its complex architecture [2].

Despite this complexity, empirical studies show that developers tend to employ only a small, common subset of M4 and shell macros, following repetitive patterns rather than arbitrary custom constructs [2]. In the PHP `config.m4` and `configure.ac` files, this subset centers on a handful of macros—`PHP_NEW_EXTENSION`, `PHP_ADD_SOURCES`, `PHP_ARG_ENABLE`, `PHP_ARG_WITH` and related functions—that govern inclusion of C source files.

Inspired by these insights and the "approximate parsing" philosophy of AutoHaven [2], our PHP resolver implements a per-file analysis that computes the presence condition of a target C source file without executing Autotools:

(1) Determine the condition of the root subsystem where the target lies in the main `configure.ac` (*e.g.*, `ext/`, `sapi/`, `Zend`...).
(2) Select the nearest file named `config.m4`, `config0.m4`, or `configure.ac/configure.in`.

[11] Approximately ~1.14 affected files per Linux patch in our dataset.

(3) Search for `PHP_NEW_EXTENSION` or `PHP_ADD_SOURCES` statements referencing the target file.
(4) Identify the surrounding `if test ... fi` blocks that wrap these statements to collect guard variables (*e.g.*, `PHP_MYEXT`).
(5) Map each guard variable to its defining macro (`PHP_ARG_ENABLE` or `PHP_ARG_WITH`) to extract the corresponding `--enable-` or `--with-` flag and its default.
(6) Include additional conditions from dependencies declared via `PHP_ADD_EXTENSION_DEP`.
(7) Conjoin all flags to form the file's overall presence condition.

Like AutoHaven [2], we rely on conservative over-approximation: if a source file is not referenced through the expected macros, our tool reports failure rather than producing an unsound condition. Resolver success rates are detailed in Section 4.

*Build system analysis for FFmpeg.* FFmpeg is a widely used open-source multimedia framework providing libraries and command-line tools for audio and video processing. Its build system consists of a hand-written shell `configure` script that probes the host environment and sets feature variables, followed by recursive GNU Make invocations across component directories (e.g. `libavcodec`, `libavformat`, `fftools`). Variability is expressed via parameters passed to the `configure` script (*e.g.*, `-enable-foo`, `-disable-bar`), which set configuration options for the build (*e.g.*, `CONFIG_FOO=yes`), consequently driving conditional object assignments in Makefiles, as shown in Listing 3.

Listing 3. Excerpt from libavfilter/Makefile, in the FFmpeg 4.2.11 source code

```
1 [...]
2 OBJS-$(HAVE_LIBC_MSVCRT) += file_open.o
3 OBJS-$(HAVE_THREADS) += pthread.o
4
5 # subsystems
6 OBJS-$(CONFIG_QSVVPP) += qsvvpp.o
7 OBJS-$(CONFIG_SCENE_SAD) += scene_sad.o
8 OBJS-$(CONFIG_DNN) += dnn_filter_common.o
9 [...]
```

Inspired by our Linux and PHP resolvers, we employ a lightweight, per-file analysis that avoids a full build. We begin by recursively collecting all `Makefile` files in the FFmpeg tree and searching each for a small, repetitive set of patterns, *i.e.*, conditional assignments (`OBJS-$(CONFIG_X)`, `SHLIBOBJS-$(CONFIG_Y)`), program definitions (`AVPROGS-$(CONFIG_Z)`), and unconditional listings (`OBJS += ...`). These matches yield a map from each object file (*e.g.*, `foo.o`) to one or more groups of feature variables, where each group represents the conjunction of all conditions required for that object.

Next, we propagate program-level guards into their constituent objects so that, for example, all object files of the `ffmpeg` program inherit its feature conditions. For the target object, we collect its associated groups and synthesize a presence predicate by AND-combining conditions within each group and OR-combining across groups, defaulting to unconditional inclusion when no real guards appear.

Finally, we derive a component-level guard from the file's top-level directory. We manually mapped each root component to its configuration option: libraries in `libXXX/` to `CONFIG_XXX`—as evidenced by conditional Makefile lines like `OBJS-$(CONFIG_XXX) += ...`—and tools under `fftools/` to the negation of `CONFIG_DISABLE_PROGRAMS`. Since our parser only ever encounters those variables when they are actually defined and used in the build scripts, this mapping reflects the build system's own definitions.

### 3.4 Presence Condition Simplification

The raw Boolean formulas produced by *PatchLens*, while sound, can be verbose and redundant—a long-standing concern in SPL analysis [52]. We therefore treat simplification as a minimum-equivalent-expression problem [9, 23] and apply established techniques, notably `RESTRICT`, Quine–McCluskey, and `ESPRESSO` [52]. Empirically, `RESTRICT` (JavaBDD-based) often yields slightly smaller results than alternatives [52], and general-purpose SMT solvers such as *z3* can assist with equivalence checking and algebraic reductions, though their effectiveness for presence-condition minimization has not been systematically compared in HCS [11]. *PatchLens* integrates simplification into its pipeline and exposes a pluggable interface; by default, it ships adapters for `RESTRICT` and *z3*, ensuring reported predicates are both precise and succinct for downstream security workflows.

### 3.5 Soundness Tests

The workflow described above yields a large dataset of vulnerabilities and their corresponding Vulnerability Impact Conditions (VICs). Since the utility of this dataset depends on correctness, we performed two soundness tests to verify that each VIC produced by *PatchLens* captures the configuration constraints under which the vulnerable code is present in a build. The methodology is presented below and results confirming the soundness of our dataset are reported in Section 4.

**Manual Inspection.** For each system analyzed, a random sample of it's VICs were selected with sufficient size to achieve a confidence level of 95% and a 5% margin of error. In this step, Cochran's formula for finite populations was used in the VICs for each system, yielding a sample size of *500* manually inspected entries.

Each sampled entry was independently reviewed by two members of the author team. Reviewers were given the *PatchLens* report and instructed to reconstruct the VIC by (i) locating the code regions modified by the patch, (ii) extracting the guarding conditional-compilation constraints (e.g., `#if`/`#ifdef`), and (iii) deriving file-level presence conditions from the build system.

For each entry, we recorded (1) report metadata (e.g., CVE, product, commit, modified files), (2) external-tool output when available, (3) the reviewer-derived condition and each step taken to produce this result, and (4) a final verdict (*sound* if the derived condition matches the VIC up to logical equivalence, *unsound* otherwise).

**Linux Kernel.** For the Linux kernel, prior work provides automated extraction of file presence conditions from Kbuild/Makefiles [7, 12, 17]. We therefore augmented Linux reports with KMAX and PCLOCATOR output [17, 30] and used it as an independent cross-check of *PatchLens*' build-system conditions; patch locations and preprocessor guards were still verified manually.

## 4 Results

This section reports the answers to our targeted research questions and findings from evaluating *PatchLens* on three large real-world C/C++ codebases: the Linux kernel, FFmpeg, and PHP.

*RQ1: To what extent can PatchLens statically analyze real-world vulnerability patches, and what is the accuracy of the results?* To answer RQ1, we applied the *PatchLens* pipeline, including AST-based diff parsing, presence condition extraction, and build system resolution to a total of 1,192, 289, and 100 Linux, FFmpeg and PHP vulnerability patches, respectively. For each patch, we recorded whether *PatchLens* successfully computed a presence condition and, if not, the failure reason.

*PatchLens* required only, on average, **47.47** milliseconds per patch to parse the diff, traverse the AST, and compute the Boolean impact formula. The breakdown of the number of cases, success rates, and the average time per project is shown in Table 1.

Table 1. *PatchLens* results across projects.

| Project | Total | Success | Failure | Avg time (ms) |
|---|---|---|---|---|
| Linux | 1,192 | 1,057 (≈ 88.67%) | 135 (≈ 11.32%) | ≈ 47.4ms |
| FFmpeg | 289 | 280 (≈ 96.88%) | 9 (≈ 3.11%) | ≈ 86.3ms |
| PHP | 100 | 68 (68%) | 32 (32%) | ≈ 87.1ms |

Subsequently, we manually reviewed a statistically representative sample of the computed VICs following the protocol in Section 3.5. In total, we reviewed 282 Linux, 160 FFmpeg, and 58 PHP reports (500 patches overall), a sample size chosen to achieve a 95% confidence level with a 5% margin of error for the VICs related to each system in our dataset. Across all reviewed entries, we found no inaccurate VICs, providing empirical evidence that the automatically derived impact conditions correctly characterize the affected variant subspaces within our dataset. Moreover, Linux VICs were cross-checked against Kmax and PCLocator [17, 30] and presented no inconsistencies.

In a few cases, reviewers observed that *PatchLens* returned a conservative *over-approximation* (*i.e.*, a VIC that includes additional variants beyond those that are truly affected). Specifically, 27 of the 500 reviewed reports (5.4%) were identified as over-approximate; when this occurred, the imprecision was typically small, corresponding on average to ≈ 1.14 missing configuration flags in the expression. Overall, these results indicate that *PatchLens* extracts accurate VICs in practice, while occasionally trading precision for safety: over-approximation enables partial yet recall-safe conditions in the presence of parser or build-system nonstandard constructs, avoiding false negatives while still providing actionable guidance.

To illustrate the over-approximation strategy implemented by *PatchLens*, Listing 4 contains one of the few reports from our supplementary materials that contain over-approximation [44]. Specifically, in this example report, *PatchLens* computes the VIC as `CONFIG_NET_SCH_QFQ` but omits the additional requirement `CONFIG_NET`. This occurs because the top-level Linux Makefile contains constructs that *PatchLens* cannot reliably interpret, preventing extraction of the full file-level presence condition. To remain recall-safe, *PatchLens* therefore conservatively assumes that the affected subtree (e.g., `net/`) may be built unconditionally, producing a superset of the truly affected variants rather than risking false negatives. Additional examples illustrating this conservative fallback are provided in the supplementary material [44].

**Observation 1**

*PatchLens* uses static analysis to rapidly compute presence conditions for the vast majority of real-world patches without building variants—≈ 88.67% on Linux, ≈ 96.88% on FFmpeg, and 68% on PHP. Moreover, manual validation of 500 sampled patches found no incorrect VICs and only occasional conservative over-approximation (5.4%), supporting its suitability for CI and large-scale studies.

*RQ2: How many security vulnerabilities impact all variants of a statically configured software system? Does a vulnerability that affects all variants have a higher impact metric?* After processing and simplifying the impact formulas for all patches, we found that only a small fraction of vulnerabilities affect every conceivable variant. In the Linux kernel, 66 patches (≈ 6.24%) of the 1,057 analyzed exhibit a tautological presence condition, indicating system-wide vulnerabilities. In PHP, 27 of the 68 analyzed patches (≈ 39.70%) likewise affect every variant, reflecting less granular variability in core interpreter features. In contrast, none of the 280 FFmpeg patches impact every configuration.

Listing 4. Example report from our validation methodology that contains a case of over-approximation

```
# Report for CVE-2023-31436 - 3037933448f60f9acb705997eae62013ecb81e0d

## Basic information
- CVE-ID: CVE-2023-31436
- Product: linux/linux_kernel
- Patch ID: 3037933448f60f9acb705997eae62013ecb81e0d
- VIC from Patchlens: "defined(CONFIG_NET_SCH_QFQ)"

### Files modified by patch
- net/sched/sch_qfq.c

## Information from external tools
- Final flag from Kmax: "(CONFIG_NET_SCH_QFQ) & (CONFIG_NET)"

## Comments
- Patchlens reported "defined(CONFIG_NET_SCH_QFQ)" while Kmax surfaced "(CONFIG_NET_SCH_QFQ) & (
      CONFIG_NET)"; manual inspection confirms the correctness of Kmax, implying that Patchlens also
      embraces all relevant configurations.
- net/sched/Makefile:55 adds `sch_qfq.o` via `obj-$(CONFIG_NET_SCH_QFQ)`, so the file affected by the
      patch is compiled once that tristate CONFIG_NET_SCH_QFQ is `yes` or `module`.
- The parent Makefile at `net/Makefile` unconditionally adds the subfolder `net/sched/` for compilation.
- The root `Kbuild` file adds the `net/` subfolder if `CONFIG_NET` is enabled. However, Patchlens fails to
       parse this file correctly and assumes `net/` is always added (over-approximation).
- The changed code around `qfq_change_class` (net/sched/sch_qfq.c:421-438) is not wrapped in any
      additional `#ifdef`s.
- Although `CONFIG_NET` is required in order to enable `CONFIG_NET_SCH_QFQ`, Patchlens conservativelly
      stops at `CONFIG_NET_SCH_QFQ`.

## Veredict (OK,UNSOUND)
OK
```

FFmpeg's build system affords developers fine-grained control over component inclusion—core libraries such as `libavcodec` and `libavformat` can be omitted with `-disable-avcodec` and `-disable-avformat`, all user-land programs removed via `-disable-programs`, and even the `ffmpeg` binary itself excluded with `-disable-ffmpeg`. Such selective exclusion ensures that virtually no vulnerability spans the entire configuration space.

We further examined severity by comparing CVSS v2/v3 **impact scores** for system-wide CVEs (i.e., patches whose VIC is tautological) against all others. The impact score captures the *direct consequences* of successful exploitation on a system; when linked to configuration options, it therefore highlights features whose vulnerabilities tend to yield higher real-world damage. The overall mean across our dataset is $\approx 4.76$, whereas system-wide CVEs average $\approx 5.94$ —an absolute increase of 1.18 points ($\approx 24\%$ higher). To assess whether this gap is unlikely to be due to chance, we applied a Mann–Whitney $U$ test, which checks whether scores in one group tend to be larger than in the other; it yielded $p = 1.15 \times 10^{-8}$. We also computed a 10,000-replicate bootstrap confidence interval for the mean difference, $[0.73,\ 1.62]$, which does not include zero. Together, these results indicate that vulnerabilities with broader variant impact tend to receive higher CVSS impact scores in practice.

To identify which features most frequently appear in vulnerabilities classified with High (7.0–8.9) or Critical (9.0–10.0) CVSS **severity scores**, we used the CVSS base score as an overall severity grade, which combines the impact score with an *exploitability* component that reflects how easily an attack can be carried out. We applied a *minimum support* filter and retained only variables that appear in at least five impact formulas before averaging their CVSS base scores. A threshold of $k = 5$ balances coverage and stability: CVSS lies on a bounded 0–10 scale, so with $n \geq 5$ the influence of any single outlier on the mean is capped at $10/n \leq 2.0$ points, whereas with $n = 3$ a single CVE can shift the average by up to 3.33 points. The top five variables per system under this criterion are reported in Table 2; in PHP, only two variables meet the five-occurrence requirement.

This conservative threshold ensures that any reported feature–severity association is supported by at least five distinct CVE observations.

**Observation 2**

Most vulnerabilities require specific configurations to have practical impact. Furthermore, the share of patches that affect *all* variants is highly project-dependent—$\approx$ 6.24% in Linux, $\approx$ 39.70% in PHP, and 0% in FFmpeg—reflecting each project's variability design (coarser-grained core options in PHP vs. fine-grained, easily excluded components in FFmpeg). When such system-wide changes occur, they are associated with *significantly* higher severity (mean CVSS $\approx$ 5.94 vs. $\approx$ 4.76). *PatchLens* detects these cases as VICs that are always `true`, enabling maintainers to quickly locate and prioritize configuration-critical code: whether code sections that impact all variants or specific configuration options that are frequently associated with CVSS scores.

*RQ3: How do vulnerability impact conditions look like?* We quantify the complexity of the presence condition by the average number of distinct configuration variables per patch and by identifying the options that most frequently govern the inclusion of vulnerabilities. Table 3 summarizes these metrics for the Linux kernel, and Figure 4 shows the distribution of variable counts across all formulas. The low average counts indicate that most security patches require only a handful of features, suggesting that targeted mitigation, such as disabling specific options or prioritizing tests for particular feature combinations, can effectively reduce the effort required for comprehensive security assurance.

Empirically, the impact conditions closely mirror how variability is implemented in each codebase. Consider FFmpeg: the file `libavformat/img2.c` is compiled whenever *any* of 51 feature options (*e.g.*, specific muxers/demuxers) is enabled; consequently, a patch that modifies this file induces an impact condition that is at least a 51-way disjunction over those options. Furthermore, the dominant options in the impact formulas reflect the architectural concerns of each system: Linux is driven by networking and I/O subsystems (*e.g.*, `CONFIG_NET`, `CONFIG_USB`), FFmpeg by codec and container modules (*e.g.*, `CONFIG_AVCODEC`, `CONFIG_AVFORMAT`), and PHP by core extensions (*e.g.*, `PHP_LIBXML`, `PHP_WDDX`).

We also examined the distribution of logical operators within each presence condition. In the Linux kernel, formulas contain 656 conjunctions ($\land$), 294 disjunctions ($\lor$), and 19 negations ($\neg$). FFmpeg formulas feature 283 conjunctions, 341 disjunctions, and a single negation. PHP formulas exhibit minimal logical structure—29 conjunctions, 1 disjunction, and 11 negations—reflecting their low average variable count. This operator distribution underscores the relative simplicity of most impact conditions, which predominantly combine a small number of feature flags using AND and OR, with only occasional negations.

**Observation 3**

Impact conditions involve on average fewer than four variables per patch (1.84 for Linux, 3.23 for FFmpeg, 1.04 for PHP), and the top ten options align with each system's key subsystems.

*RQ4: How much information can PatchLens add to CVE descriptions?* We assessed how often CVE descriptions can be enriched with the *compile-time* options that actually gate each vulnerability. For every CVE with a computed impact condition, we extracted the ground-truth option set $G_i$

Table 2. Top 5 configuration options by average CVSS base score (per variable)

| Linux | | FFmpeg | | PHP | |
|---|---|---|---|---|---|
| **Option** | **Average CVSS** | **Option** | **Average CVSS** | **Option** | **Average CVSS** |
| CONFIG_NFSD | 8.04 | CONFIG_WMALOSSLESS_DECODER | 9.60 | PHP_LIBXML | 6.83 |
| CONFIG_SMB_SERVER | 8.00 | CONFIG_TIFF_DECODER | 8.08 | PHP_WDDX | 6.64 |
| CONFIG_NFSD_V4 | 7.83 | CONFIG_SANM_DECODER | 7.82 | – | – |
| CONFIG_IO_URING | 6.93 | CONFIG_PNG_DECODER | 7.66 | – | – |
| CONFIG_BPF | 6.91 | CONFIG_H264_PARSER | 7.60 | – | – |

Table 3. Average number of variables per impact formula and top 10 configuration options by frequency

| Linux (avg. 1.84 vars) | | FFmpeg (avg. 3.23 vars) | | PHP (avg. 1.04 vars) | |
|---|---|---|---|---|---|
| **Option** | **Count** | **Option** | **Count** | **Option** | **Count** |
| CONFIG_NET | 77 | CONFIG_AVCODEC | 210 | PHP_LIBXML | 12 |
| CONFIG_INET | 51 | CONFIG_AVFORMAT | 52 | PHP_WDDX | 9 |
| (ARCH = x86) | 48 | CONFIG_AVFILTER | 21 | HAVE_WDDX | 9 |
| CONFIG_USB | 45 | CONFIG_JPEG2000_DECODER | 12 | (*PHP_GD* = "no") | 4 |
| CONFIG_USB_GADGET | 42 | CONFIG_H264_DECODER | 12 | PHP_PHAR | 3 |
| CONFIG_PCI | 38 | CONFIG_SVQ3_DECODER | 11 | PHP_ZIP | 3 |
| CONFIG_SND | 37 | CONFIG_H263_DECODER | 10 | PHP_INTL | 3 |
| CONFIG_KVM | 35 | CONFIG_MPEG4VIDEO_PARSER | 10 | ZTS | 3 |
| CONFIG_USB_PHY | 32 | CONFIG_H264_PARSER | 10 | PHP_SOAP | 2 |
| CONFIG_NETFILTER | 30 | CONFIG_PNG_DECODER | 8 | (*PHP_SNMP* = "no") | 2 |

(*e.g.*, `CONFIG_KVM`, ARCH = x86), tokenized the NVD description, and searched for mentions of these options. Because option names in our corpus are typically uppercase, delimiter-rich identifiers (e.g., `CONFIG_AVCODEC`, `PHP_LIBXML`), we used a conservative fuzzy match to tolerate minor formatting variations (underscore vs. hyphen, case, singular/plural) while avoiding spurious hits. Concretely, we required a word-boundary match and a `difflib` similarity of $\geq 0.8$. This threshold is intentionally high: for medium-length identifiers (8–15 characters), a single edit (e.g., `CONFIG-KVM` vs. `CONFIG_KVM`) still exceeds 0.8, but generic tokens with incidental overlap (e.g., "net", "usb") fall well below it.

Applying this pipeline to all 1,405 CVEs successfully processed by *PatchLens*, we find that only **1.50%** of descriptions mention *at least one* required option (coverage), and on average only **0.90%** of the options present in an impact formula are mentioned (recall). These results indicate substantial headroom for automated enrichment: inserting the precise, human-readable VIC computed by *PatchLens* would materially improve the practical utility of CVE entries for maintainers and operators.

**Observation 4**

Only **0.90%** of configuration options required for the vulnerability to be present are mentioned in CVE descriptions, revealing the critical need for automated enrichment with precise VICs.

## 5 Discussion

In this section, we discuss the broader implications of *PatchLens* and the associated patch impact condition data set, highlighting how these contributions can drive advances in both industrial practice and academic research.

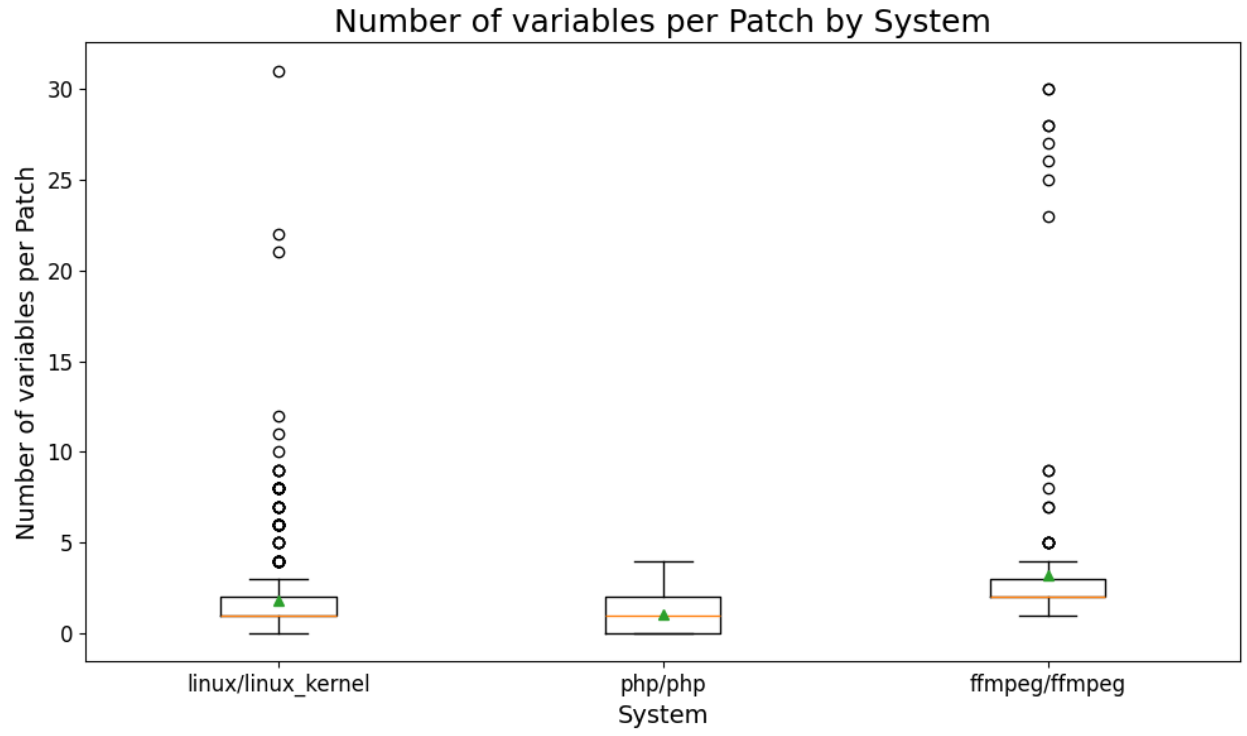

Fig. 4. Distribution of number of configuration variables per patch by system.

## 5.1 Implications for Industry

*PatchLens* 's precise, human-readable Vulnerability Impact Conditions (VICs) can be directly integrated into vulnerability management workflows. With the support of *PatchLens*, vulnerability databases can add the conditions that gate each vulnerability. This additional information allows operators and maintainers to rapidly determine whether their deployed variants are affected and, consequently, avoid unnecessary patch roll-outs and conserve resources [32]. For example, Figure 5 illustrates a CVE entry with this additional information generated by *PatchLens*. Furthermore, *PatchLens* 's fast processing time (≈ 47ms) enables seamless insertion into CI/CD pipelines or into IDEs, providing immediate feedback on the variant scope of code changes without full builds or checkouts.

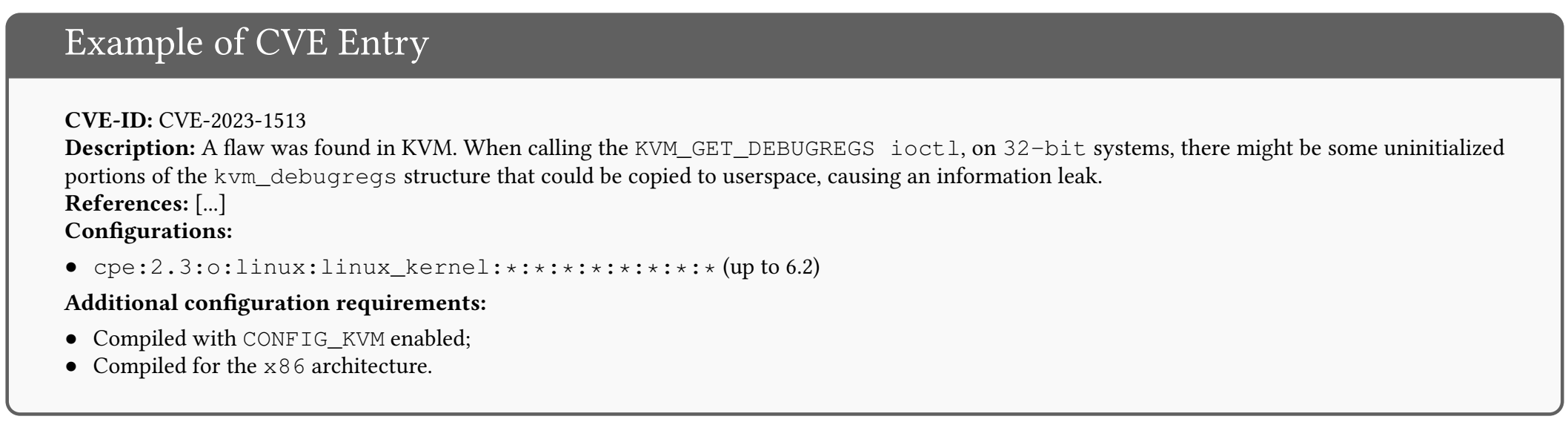

Example of CVE Entry

**CVE-ID:** CVE-2023-1513
**Description:** A flaw was found in KVM. When calling the `KVM_GET_DEBUGREGS ioctl`, on `32-bit` systems, there might be some uninitialized portions of the `kvm_debugregs` structure that could be copied to userspace, causing an information leak.
**References:** [...]
**Configurations:**

- `cpe:2.3:o:linux:linux_kernel:*:*:*:*:*:*:*:*` (up to 6.2)

**Additional configuration requirements:**

- Compiled with `CONFIG_KVM` enabled;
- Compiled for the `x86` architecture.

Fig. 5. Example of a CVE Entry with additional configuration requirements generated by *PatchLens*

It is important to clarify that *PatchLens* is designed as a unified framework with an adapter architecture for build-system analysis: we currently provide initial implementations for the Linux kernel, PHP, and FFmpeg. However, we expect that extending support to additional systems would typically require only ≈200–400 lines of code to map target definitions to file paths (or fewer when adapting an existing resolver).

Beyond maintenance, *PatchLens* can improve software development processes such as regressive test selection. Regression testing seeks to re-run only those unit tests impacted by a change—a task that can be optimized by filtering test suites based on patch impact formulas [48]. By mapping each test's coverage to the affected impact conditions, teams can execute only relevant tests, dramatically reducing turnaround time.

Moreover, the rich dataset of patch impact conditions enables a new form of feature-risk scoring. By aggregating the frequency and severity of vulnerabilities tied to each compile-time flag, maintainers can compute a risk rank for every option and adopt *secure-by-default* build profiles that disable the highest-risk features. This data-driven hardening minimizes attack surface from the outset. Finally, the same dataset can enrich software bill of materials (SBOM) artifacts: annotating each component with the set of configuration flags historically implicated in CVEs allows auditors and customers to quickly determine which features warrant additional scrutiny or remediation.

### 5.2 Implications for Academia

In the research area of variability-aware testing, *PatchLens* provides a concrete basis for *targeted* combinatorial interaction testing (CIT) and sampling strategies [22, 25, 37]. Rather than optimizing for uniform coverage over the entire configuration space, researchers can explicitly focus on the *variant subspaces* delineated by impact formulas. Empirically, from **RQ3**, we observe that the average number of configuration options appearing in a VIC is small and exhibits low variance (see Figure 4). This suggests a practical design heuristic: sampling strategies that preferentially generate variants whose interaction strength matches the empirical VIC complexity of the system may achieve superior fault-revelation per build. For instance, the Linux kernel shows an average of 1.84 options per VIC; consequently, 2-wise sampling is a strong candidate baseline for Linux-like systems, as it aligns the sampling strength with the observed structural complexity of security-relevant changes.

Beyond CIT, *PatchLens*-derived datasets can seed guided fuzzing and test-case prioritization. Historical predicates reveal *error-prone subspaces*—combinations of features that recurrently gate vulnerabilities—which enable fuzzers and search-based generators to allocate effort where returns are highest (cf. Table 3 and Table 2 for representative high-risk subspaces). Likewise, the corpus of VICs forms a realistic benchmark to evaluate t-wise and covering-array strategies against *actual* vulnerability distributions, ensuring that methodological innovations are validated on security-relevant targets rather than synthetic surrogates.

The same representation supports learning-based studies: encoding each VIC as a multi-label feature vector and pairing it with patch metadata (e.g., CVSS, changed subsystems) yields a natural training set to predict both configuration scope and severity, guiding proactive triage before costly static or dynamic analyses. Longitudinally, trends in VIC structure can reveal emerging hotspots, inform feature deprecation, and refine variability-management strategies. By grounding these investigations in precise, patch-derived predicates, *PatchLens* links methodological insights to measurable outcomes in large C/C++ highly configurable systems.

## 6 Threats to Validity

**Construct Validity**: Macro and preprocessor complexity, such as deeply nested conditional directives, code generation via macros, and nonstandard preprocessor extensions, can impede accurate AST-based extraction of VICs. This limitation, exemplified by PHP's 32% processing failure rate, may lead to incomplete or imprecise impact formulas. However, only a small portion of patches in our dataset required parsing such complex constructs, approximately 11%—which limits the practical impact of this threat.

**Internal Validity**: *PatchLens* does not model the transitive effect of changes in header files on including translation units. Thus, vulnerabilities introduced or remediated in the headers can be under-reported, biasing our estimates of affected variants. Additionally, our patch corpus is drawn solely from GitHub commit references in NVD feeds and from PatchDB. Vulnerabilities fixed through other channels—such as proprietary patches, fixes on mailing lists, or vendor advisories without GitHub commits—are excluded, potentially skewing both the project distribution and the types of vulnerabilities analyzed. Nevertheless, our datasets representativeness is strengthened by PatchDB's construction, which blends NVD-indexed and in-the-wild patches—12K security fixes from 313 GitHub repositories, manually cross-checked by security experts [53]. Additionally, in our corpus, most security patches modify implementation files rather than headers, so the practical impact of this limitation is modest.

**External Validity**: All experiments target C and C++ codebases. Systems implemented in languages with runtime or dynamic variability (*e.g.*, Java, Python, JavaScript), or mixed-language projects, may employ different configuration mechanisms and are not covered by our evaluation. However, a substantial share of highly configurable large-scale systems — and much of the literature on HCS – is concentrated in C/C++ codebases, where compile-time variability via the preprocessor and build systems is pervasive; thus, our scope aligns with a widely studied and practically consequential segment of the ecosystem [1, 15, 27, 36, 50, 52].

## 7 Related Work

**Localizing configurations in HCS**: Variability-aware techniques analyze all program variants as a single artifact rather than one configuration at a time [1, 15, 27, 36, 50, 52]. Due to heavy computational requirements, a central challenge is finding which valid variants of a system include a specific portion of the code [19]. Specialized parsers such as TypeChef and SuperC address this by producing representations (*e.g.*, ASTs with choice nodes) that embed compile-time variability [15, 18, 27]. Building on these foundations, researchers have proposed variability-aware type checking, data-flow analysis, and model checking, though scaling to large systems remains difficult; simplifying presence conditions is therefore essential for tractability [52]. Closely related, PCLocator automatically identifies configurations that include a given code location by extracting and composing presence conditions from source and build artifacts, enabling solver-backed localization without exhaustively enumerating variants [30].

In this context, *PatchLens* adopts the same premise of operating on unpreprocessed code but targets a different unit of analysis: it computes *patch-scoped* impact formulas by mapping diff hunks to their variability contexts and resolving build system predicates, then simplifies the resulting expressions into human-readable VICs [52]. Compared with SiB ("Should I Bother?"), which safely filters *known, pre-built* variants by intersecting patch hunks with a line-range database recorded during compilation—thus requiring at least one build per variant [32]—*PatchLens* derives a *global* Boolean predicate over configuration options without building any variant. The approaches are complementary: SiB offers fast, recall-safe fleet filtering for shipped configurations, while *PatchLens* enables explainable, configuration-wide reasoning and documentation of patch impact.

Furthermore, to determine the full presence condition of a code artifact inside a highly configurable system, tools have to inspect both the source code for variability-related macros and the build system itself [19]. However, build system analysis is non-trivial due to the non-standard nature of these. Consequently, an universal tool for build system presence condition extraction is unavailable.

**Build System Variability**: Build systems such as Kbuild and GNU Autotools do not merely run compilation commands; they implement variability by selecting files and options for concrete variants [2, 12, 50]. Because these specifications span multiple languages (*e.g.*, Make, shell, M4), macro expansion, and multi-stage processing, extracting reliable presence conditions from build logic is non-trivial and often defeats simple text parsing [2, 12]. Prior work has therefore developed dedicated, semantics-aware analyzers for Kbuild: Kmax statically computes file-level presence conditions from Kbuild Makefiles across the configuration space [17]. *PatchLens* makes build system variability first-class in computing Vulnerability Impact Conditions (VICs): it isolates source-level guards, inserts explicit `building()` placeholders, and resolves them via project-specific analyzers (*e.g.*, Linux/Kbuild, FFmpeg/Make, PHP/Autotools).

**Bugs, Faults, and Maintenance in Configurable Systems**: Configurable systems exhibit faults that manifest only under specific combination of options: variableness and interaction bugs of features - documented in large codebases such as the Linux kernel [1, 16, 24, 37]. These arise both in C code (e.g., interactions across conditional regions) and in configuration/assembly logic, including build-system inconsistencies [1, 12, 16, 24, 37, 51]. Preprocessor pitfalls further show how tangled annotations introduce syntax and maintenance hazards [39], while human and process factors contribute to vulnerability-inducing commits [24]. The resulting maintenance burden—especially in build code—has motivated variability-aware refactorings and analyses [2, 12, 14, 15, 32, 34]. Recent work has also leveraged these variability constraints to improve testing: KREPAIR generates a small set of configurations that maximize patch compilation coverage while remaining close to a baseline configuration [54]. *PatchLens* complements this body of work by translating each security patch into a human-readable VIC that (i) localizes feature-interaction risk by naming the option combinations implicated in a fix [1, 16, 37], (ii) incorporates build semantics to recover file-level presence conditions [2, 12], and (iii) provides a reusable predicate that improves triage, documentation, and regression testing across the configuration space [15, 32, 34].

## 8 Conclusion and Future Work

In this paper, we introduced *PatchLens*, a purely static analysis tool designed to precisely characterize the set of software variants affected by security patches through Boolean presence conditions. By operating directly on AST-level differences and carefully integrating variability constraints from the build system, *PatchLens* provides sound and explainable patch-impact analysis without the overhead of compiling multiple variants. Our empirical evaluation, conducted across 1,192 Linux kernel patches, 289 FFmpeg patches, and 100 PHP patches, demonstrates that *PatchLens* rapidly computes impact formulas. Additionally, we found that vulnerabilities generally depend on very few configuration options and that this information can be added to the metadata of many vulnerabilities, *e.g.*, to enrich CVE descriptions.

As future work, we intend to extend *PatchLens* in several directions. First, we aim to incorporate configuration data from widely used software distribution repositories, *e.g.*, Debian and Fedora, allowing us to assess how they deviate from default settings. Additionally, we plan to enhance *PatchLens* to fully account for header-file impacts by analyzing include dependencies. Finally, we believe that expanding our CVE dataset to encompass a broader range of software ecosystems will also validate and extend the applicability of our tool.

## 9 Data Availability

An anonymized replication package containing all source code, data from experiments, analysis scripts, and instructions on how to replicate this study has been provided as supplemental material for review. [44]. Additionally, we provide a GitHub repository for future updates in our implementation [43].

## Acknowledgments

This work was partially supported by INES.IA (National Institute of Science and Technology for Software Engineering Based on and for Artificial Intelligence), www.ines.org.br, CNPq grant 408817/2024-0, CNPq grant 444956/2024-7, and FAPESB INCITE PIE0002/2022 grant. This work was partially supported by CAPES and FAPERJ under grants E-26/204.268/2024 and E-26/260.168/2026, CNPq grants 444956/2024-7, 424622/2021-1 and 315106/2023-9, as well as Finep PlatCiber.